\documentstyle[prl,aps,floats,preprint]{revtex}   
\begin{document}
\draft


\def\rg2{{\big<R_G^2 \big>}}
\def\r2{{\big<R^2\big>}}
\def\bull{{\vrule height .9ex width .8ex depth -.1ex }} 


\title{Dynamics of $n$-alkanes: Comparison to Rouse Model}

\author{Maurizio Mondello, Gary S.\ Grest, Edmund B.\ Webb~III, P.\ Peczak,
and  Scott T.\ Milner}

\address{Corporate Research Science Laboratories,
Exxon Research and Engineering Company,
Annandale, New Jersey 08801}

\date{\today}

\maketitle
{
\begin{abstract}
The crossover to Rouse-like behavior  for the self-diffusion 
constant $D$, the viscosity $\eta$, and the equilibrium structural
statistics
of $n$-alkanes $(6 \le n \le 66)$ is studied numerically. 
For small $n$ the chains are non-Gaussian and the mean
squared end-to-end distance $\r2$ is greater than $6\rg2$, where
$\rg2$ is the mean squared radius of gyration. As $n$ increases, $\r2/\rg2 
\rightarrow 6(1+a/n)$, where $a$ depends on the interaction model.
At constant density, the Rouse model is used to extract
the monomeric friction coefficient 
$\zeta$ and the viscosity $\eta$ independently from the diffusion
constant $D$ and the longest relaxation time $\tau_R$.  $\zeta_D$ extracted
from $D$ is nearly independent of chain length while 
$\zeta_\tau$ obtained from $\tau_R$ is much larger than $\zeta_D$ for 
small $n$.  The viscosity measured in a non-equilibrium
molecular dynamics simulation is closely approximated by
the value of $\eta$ determined from $\tau_R$ while 
$\eta$ inferred  from $D$ is smaller for small $n$. 
For $n\agt 60$, the two estimates for both  $\zeta$ and $\eta$ agree
as predicted from the Rouse model.
$D$ calculated from three interaction models
is studied for increasing $n$ and compared to experimental data.

\end{abstract}
}
{
\pacs{\mbox{ }}
}

\narrowtext

\noindent{\bf I. INTRODUCTION}

It is well known that dynamics of a melt of unentangled polymer chains 
satisfies the Rouse 
model \cite{rouse53}
exceedingly well.  For chains which are shorter than an entanglement
length $n_e$,  the self-diffusion constant and viscosity for a melt of 
chains of length $n$ are observed 
experimentally \cite{berry68,ferry80}
to scale as  $n^{-1}$ and $n$, respectively, as predicted by the 
Rouse model.  
Computer simulations  on simple bead spring models \cite{kremer90}
have shown that not only does the Rouse model predict the correct scaling
dependence on chain length $n$, it also gives quantitatively correct results.
Why this simple model, which  was first proposed \cite{rouse53}
to model the dynamics of dilute polymer solutions, works so well for
a polymer melt  remains somewhat of  a mystery.  

The crossover from Rouse-like behavior for
unentangled short chains to reptation-like motion for
entangled chains has received considerable attention in recent
years, while  the crossover from the behavior of very small molecules
to the Rouse regime has received relatively little attention. 
Understanding the latter crossover can provide information on
subtleties of the Rouse model and at the same time increase
knowledge on the dynamics of a technologically important class of
molecules, normal alkanes.  There
have been  a number of studies of the dynamics of short alkanes,
but to our knowledge there have been no systematic studies 
which examine the dependence of  either the diffusion or viscosity
on chain length.  One can also examine the crossover
of the mean squared radius of
gyration $\rg2$ and
end-to-end distance $\r2$ as a function of $n$.  For Gaussian chains,
$\rg2$ and $\r2$ satisfy
\begin{equation}
\label{gaussian}
\r2=6\rg2=nb^2
\end{equation}
where $b$ is the effective bond length.
Consider the simple case of normal alkanes,
though the same is true for any flexible chain molecule:  for small
$n$, the chains are clearly non-Gaussian and $\r2 > 6\rg2$.
For normal $n$-alkanes,
crossover to Gaussian chain statistics  occurs for $n$
greater than  $100$ \cite{baschnagel92,brown9496,paul98}. 
Since the Rouse model is based on the
fact that the chains are Gaussian, one would expect that crossover
length is
a minimum chain length for the Rouse model to hold, though this
has not previously been checked systematically.

In the Rouse model, the excluded volume interactions and the
hydrodynamic interactions are disregarded and the polymer is
treated as a collection of beads connected with a harmonic
spring with spring constant $k=3k_BT/b^2$. Here $T$ is
the temperature and $k_B$ is Boltzmann's constant. The dynamics of
the chain is then modeled with a Langevin equation with a monomeric
friction coefficient $\zeta$. The resulting set of 
equations can be solved by transforming to normal coordinates. The resulting
self-diffusion constant is \cite{doi86} 
\begin{equation}
D={k_BT\over n\zeta}.
\label{eq:D}
\end{equation}
The rotational relaxation time $\tau_R$ of a polymer is usually
defined by the longest relaxation time of the end-to-end
autocorrelation function  
$\big<{\bf  R}(t)\cdot {\bf  R}(0)\big>$ and expressed as \cite{doi86},
\begin{equation}
\tau_R={\zeta n^2b^2\over 3\pi^2k_BT}.
\label{eq:taur}
\end{equation}
The effective bond length $b$ can be eliminated in favor of either
$\r2$ or $\rg2$ 
using eq.~\ref{gaussian},  
\begin{equation}
\label{eq:taurrg}
\tau_R={\zeta n \r2\over 3\pi^2k_BT}={2\zeta n \rg2\over \pi^2 k_BT}.
\end{equation}
The viscosity $\eta$ of a melt can be determined from the
relaxation modulus $G(t)$, which can be written as a sum over
the Rouse modes of a chain. In terms of
$\tau_R$ \cite{doi86}, 
\begin{equation}
\label{eq:vistau}
\eta= {\pi^2\rho RT\tau_R\over 12 M},
\end{equation}
where $M$ is the molecular weight, $\rho$ is the 
density, and $R$ is the gas constant. This gives the well known
result that for Rouse chains, the viscosity scales linearly with
$n$ for constant monomeric friction coefficient. This simple scaling does not
hold at constant temperature and pressure, since $\rho$ and therefore
$\zeta$ increase with increasing $n$. This results in a stronger
dependence on $n$, namely $\eta \sim D^{-1}\sim n^{-1.8}$ for
unentangled polyethylene chains for $T=448$~K \cite{pearson87a}.

To study the crossover from small molecule behavior to Rouse-like
chains,  it is convenient to rewrite the monomeric friction coefficient
$\zeta$
and viscosity $\eta$ in several different forms. From eq.~\ref{eq:D}
for the diffusion constant
$D$, we can write
\begin{equation}
\label{eq:zetaD}
\zeta_D={k_BT\over nD}.
\end{equation}
{From} eq.~\ref{eq:taurrg} for the rotational relaxation time $\tau_R$,
we have two measures
of $\zeta$ depending on whether we use $\r2$ or $\rg2$ to eliminate
$b$,    
\begin{eqnarray}
\label{eq:zetatau}
\zeta_\tau(R)= {3\pi^2 k_BT\tau_R\over n\r2} \nonumber\\
& & \\ 
\zeta_\tau(R_G)= {\pi^2 k_BT\tau_R\over 2n\rg2}. \nonumber
\end{eqnarray}
The viscosity can also be written
three different ways. The first, given by eq.~\ref{eq:vistau}, we refer
to as $\eta_\tau$. Two additional expressions for $\eta$ in terms
of $D$, can
be obtained by substituting each form of eq.~\ref{eq:taurrg}
for $\tau_R$ in
eq.~\ref{eq:vistau} and eliminating $\zeta$ in favor of $D$ using
eq.~\ref{eq:D},
\begin{eqnarray}
\label{eq:visD}
\eta_D(R)={\rho RT\r2\over 36MD} \nonumber\\
&  \\
\eta_D(R_G)={\rho RT\rg2\over 6MD}. \nonumber 
\end{eqnarray}
These three expressions for both $\zeta$
and $\eta$ are equivalent for Rouse chains,
so that any differences are a clear signature of 
deviations from Rouse behavior.  
 
To investigate  the crossover to Rouse-like chains with increasing
chain length, we carried out a series of equilibrium (EMD) and non-equilibrium
molecular dynamics (NEMD) simulations for normal alkanes  with 
$6\le n\le 66$ at 
the
density for
high molecular weight polyethylene.
Constant density was studied due to
the large corrections to Rouse behavior observed experimentally for
unentangled chains below the entanglement length
$n_e$ at atmospheric pressure.
The EMD simulations were used to calculate both $\zeta$
and $\eta$ as predicted by the three expressions for each discussed
above.  The results for $\eta$, as predicted by the EMD calculations,
were compared to the viscosity calculated directly from the NEMD simulations.
We also carried out EMD simulations on alkanes in this size range
employing three different interaction models 
discussed recently in the literature.
These were performed
at experimental density for atmospheric pressure and the diffusion constant
$D$ as a function of 
chain length was compared to experiment.
All sets of EMD simulations were
used to study the behavior of the equilibrium
structural statistics $\rg2$ and $\r2$ with increasing chain length.
To efficiently extend this comparison out to sizes larger than
$n=66$, we also carried out simulations using the rotational isomeric
state (RIS) model for chains in the range 
$34\le n \le 2000$.

In Sec.\ II we summarize the simulation models and the
methodology used.  The equilibrium results for the static structural
properties and their behavior with increasing $n$
are presented in Sec.\ III.  Results for the dynamics
are presented in Sec.\ IV where we discuss the crossover for the 
monomeric friction coefficient and viscosity as a function of chain length
$n$ and compare diffusion results to experiment and predictions
based on the Rouse model. 
We briefly summarize our main conclusions in Sec.\  V.

\noindent{\bf II. SIMULATION MODEL AND METHOLOGY}

In order  to study a wide range of  chain lengths, an united
atom (UA) model was used to simulate the linear alkanes. In this
model, the hydrogen atoms are not
explicitly simulated.  Rather, they are grouped with the carbon to which
they are bonded into a single particle or a united atom.
We do, however, distinguish between different  ${\rm CH}_n$ groups.
These groups or united atoms interact through bonded and non-bonded forces.
The bonded interactions are represented by constraint forces which keep 
intramolecular nearest-neighbors at a fixed distance, a bending term
\begin{equation}
V_b(\theta)={k_b\over 2} (\theta-\theta_b)^2
\label{bend}
\end{equation}
where $\theta_b$ is the equilibrium angle between successive bonds, and
a torsional term
\begin{equation}
V_t(\phi)=\sum_i  a_i  \cos^i(\phi)
\label{torsional}
\end{equation}
characterizing preferred orientations and rotational barriers around all
non-terminal bonds. The non-bonded forces are described by Lennard-Jones
(LJ) interaction sites located at the position of each carbon atom center of
mass. The LJ potential is defined by 
\begin{equation}
V_{LJ} (r) = 4\epsilon\big[ \big({\sigma\over r}\big)^{12} -\big({\sigma\over r}\big)^6\big]
\label{lj}
\end{equation}
This non-bonded interaction is between both intermolecular and
intramolecular sites except for those which are separated by less
than four bonds and which therefore interact through one or more of the
bonded interaction terms. The LJ interaction is usually truncated at a
distance $r_c$ and the potential shifted so that $V_{LJ}(r_c)=0$.

In our earlier studies of the diffusion \cite{mondello95,mondello96}
and viscosity \cite{mondello97} of linear and branched alkanes, we studied
two models for the interactions. The first was the symmetric UA
model of Siepmann {\it et al.} \cite{siepmann93,smit95} which
was developed to describe the vapor-liquid coexistence curves of
the $n$-alkane phase
diagram \cite{ednote1}.
We refer to this as model A.
The second, which we denote as model B,  
was the asymmetric united atom (AUA) model of Padilla
and Toxvaerd \cite{padilla91,padilla91a} which was optimized to describe
the static
and dynamic behavior of short $n$-alkanes $(5\le n \le 10)$ at moderately
high temperatures and pressures. 
In this latter model, there is a displacement between
the centers of force of non-bonded interactions and the centers of mass of the
united atoms, hence the name.
Paul {\it et al.} \cite{paul95} recently introduced a new,
optimized symmetric UA model which they claimed
was better able to reproduce the properties of melts of linear normal alkanes.
The primary differences between their model and that of Siepmann {\it et al.}
\cite{siepmann93,smit95}
was the use of a slightly larger $\sigma$ and a different torsional potential;
we refer to our implementation of this last
parameterization as model C \cite{ednote3}.

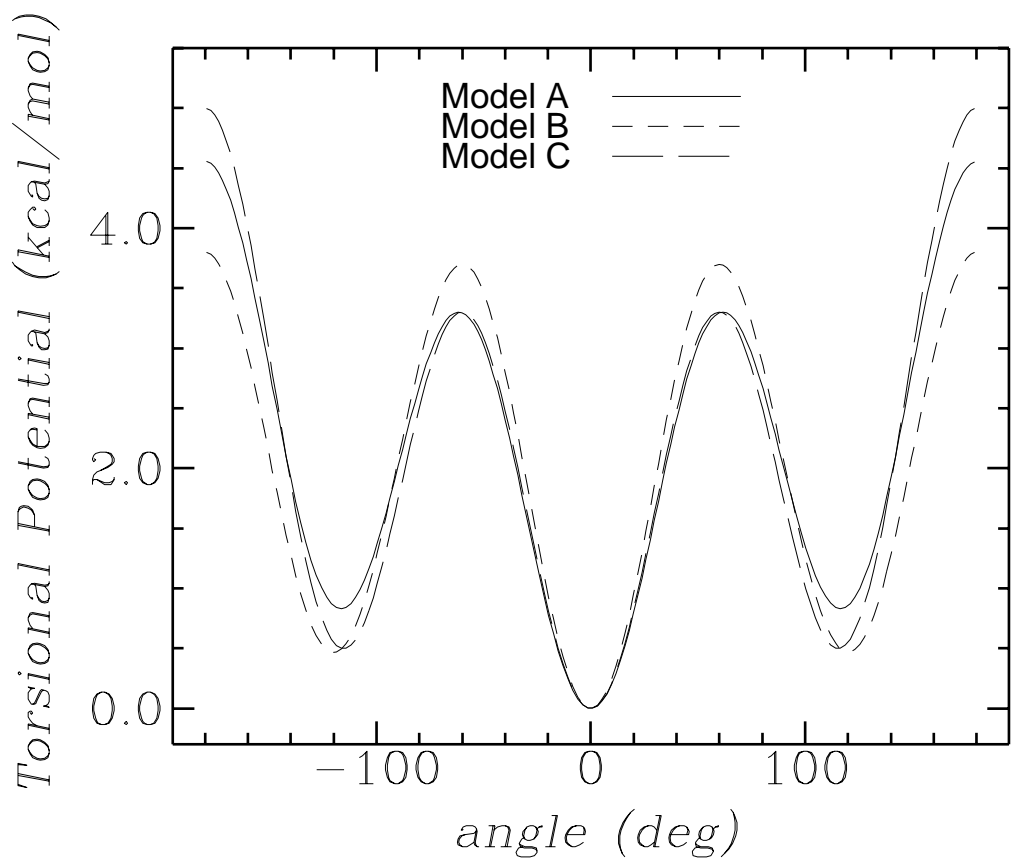
\begin{figure}[tb]
\caption{Comparison of torsional potential around $X-CH_2-CH_2-Y$ type
bonds for models A, B, and C.
\label{f:torsional}}
\end{figure}

The intramolecular interaction parameters for the three models 
are listed in Table \ref{table1}. The parameters
for the torsional potential of model A are from Jorgensen {\it et al.} 
\cite{jorgensen84}.
Paul {\it et al.} \cite{paul95} presented results for two torsional potentials.
The first one they studied
gave a  diffusion constant which was $45\%$ larger than the experimental
result for $n-C_{44}H_{90}$ at $T=400$~K.
They subsequently 
modified the
torsional potential $V_t(\phi)$ to increase 
the {\it trans-gauche} barrier from 
$3.0$ to $3.3$ kcal/mol. Since this latter model gave a better estimate of the 
$D$, though still $30\%$ too large for $n-C_{44}H_{90}$, we used their modified
version of  $V_t(\phi)$ here. For comparison, the torsional potentials
for the three models are shown in Fig.~\ref{f:torsional}. Note that
the {\it gauche}
energy is about the same for models B and C $(0.5$ kcal/mol) and about
half that of model A $(0.9$ kcal/mol). This difference will
influence the thermodynamic flexibility of the molecules and the persistence
length for long chains.  Models B and C are expected to result in
more thermodynamically
flexible molecules than model A.  The dynamic flexibility is related to the 
size of  the {\it trans-gauche} barrier, which is larger for model B
than for models A and C.  Thus, models A and C lead to more dynamically
flexible molecules than model B.

The Lennard-Jones parameters are listed in Table \ref{table2} for all three
models. Note that the interaction strength $\epsilon$ is the same for
models A and C. The effective diameter for the united atoms is
different for the
three models, with model C having a larger effective diameter than
model A.  For model B, a direct comparison based on the value of $\sigma$
is not appropriate due to the asymmetric nature of united atoms.  While
the $\sigma$ for model B is the smallest, the displacement between
the center of force of non-bonded interactions and the center of mass
of the united atom, results in an effectively larger united atom than
represented by $\sigma$.
The  interaction parameter between  a ${\rm CH}_3$ and ${\rm CH}_2$ site is
given by the  Lorenz-Bertholot rule, 
$\sqrt{\epsilon_{{\rm CH}_3}\epsilon_{{\rm CH}_2}}$. 
We used a $10\ \AA$ cutoff for the (shifted) LJ potential.

All of the MD results presented here were obtained from constant
volume, constant temperature simulations.
The velocity rescaling algorithm of Berendsen {\it et al.}
\cite{berensen84} was used to control the temperature in the EMD
simulations while the Nose-Hoover \cite{allen87}
algorithm was used for the NEMD simulations.
The equations of
motion were integrated using the velocity Verlet algorithm \cite{allen87}
with a $5$ fs time step for both the EMD and NEMD simulations.
Bond lengths were kept constant using the 
RATTLE algorithm \cite{allen87,andersen83}.
Equilibrium structural statistics, the self-diffusion
constant $D$, and rotational diffusion time $\tau_R$ were determined from
EMD simulations while the viscosity
$\eta$ was determined from
NEMD simulations. The NEMD results reported here were obtained in
the Newtonian
regime, for a shear rate ${\dot\gamma}=8.70 \times 10^{-4}\ {\rm ps}^{-1}$,
which satisfies
${\dot\gamma} < \tau_R^{-1}$ for all $n$ studied.

One set of EMD simulations was done at constant density
for a series of chain lengths $n$.  The density used was
the bulk density for high
molecular weight polyethylene,
$\rho=0.766\ {\rm g/cm}^3$ \cite{pearson87a} and the
simulations were carried out at ($T=448$~K).  Selected chain
lengths at this state point were also studied with NEMD simulations.
For small $n$, this state point corresponds to a high pressure state.
In all of the constant density simulations (EMD and NEMD),
interaction model A alone was employed.
Another set of EMD simulations were performed using the
experimental densities at atmospheric pressure $(P\simeq 0.1$~MPa)
for four chain lengths $n$.  For each of these chain lengths, simulations
were run using all three interaction models A, B, and C. These
experimental density simulations were run at $T=400$~K in
order to remain far below the vapor-liquid critical temperature
$T_c$ for small $n$, which for example is
$507$~K for $n$-hexane and $617.5$~K for $n$-decane \cite{anselme90}.
Since we were interested in studying the effect of chain length on
diffusion and viscosity,
we chose temperatures which were higher than the melting
temperature $T_m$ for large $n$. For linear polyethylene,
$T_m\simeq 400$~K.
The lengths of the runs and the number of molecules used for 
each case are presented in Table \ref{table3} and Table \ref{table4}.

To examine equilibrium structural statistics for $n > 66$, we used 
a continuum version of the RIS model for chains up to length $n=2000$.
There have been
several studies \cite{baschnagel92,brown9496,paul98,winkler91,sariban92}
which have shown that by  proper choice of the cutoff for the non-bonded
interactions, the radius of gyration and end-to-end
distance of an isolated chain compare very favorably with that of
the  melt.  Here we followed
the standard RIS model for polyethylene,
which includes all bonded interactions and the 1-5 intramolecular
Lennard-Jones potential (these are the ``second-order
interactions" of RIS theory and are known to be essential in
order to reproduce the experimental characteristic ratio of 
the polyethylene chain). The main difference between the present
model and the standard RIS model is that in the
present implementation the molecule can explore
the entire torsional potential surface, rather than being confined to a
set of its minima.  To allow comparison with the constant density
EMD simulations, the 
simulations were done at $T=448$~K. 
For short chains $(n\alt 100)$, one can use Langevin dynamics simulations
in which each isolated chain is coupled weakly to a heat bath. However
for large $n$, the pivot algorithm \cite{pivot} is considerably more
efficient. Most of the results for the RIS simulation were  obtained using
the pivot algorithm, though for small $n$ both methods gave comparable
results as expected.

\noindent{\bf III. EQUILIBRIUM PROPERTIES}

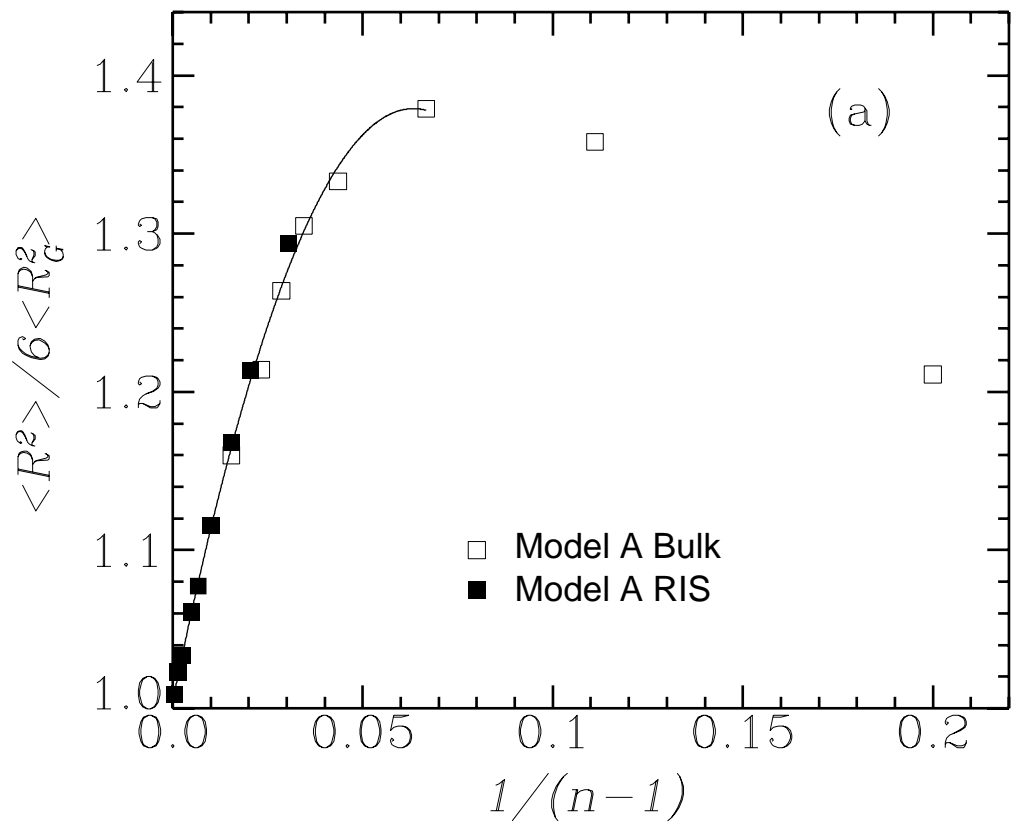
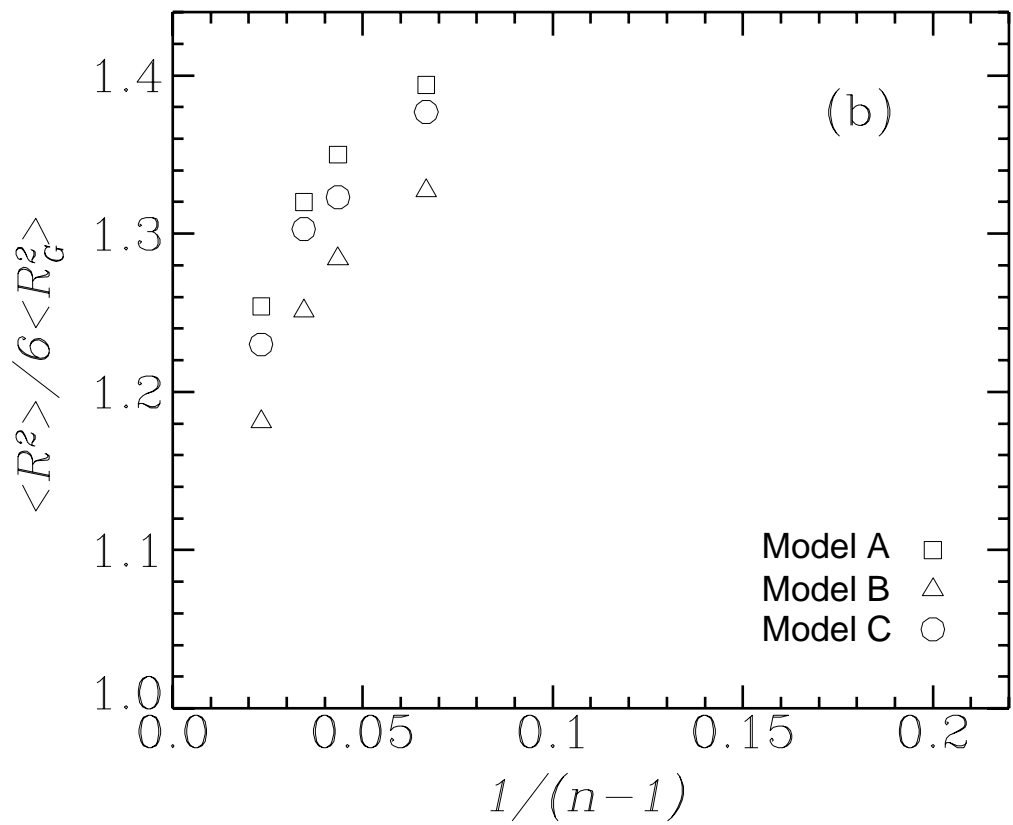
\begin{figure}[tb]
\caption{Mean squared end-to-end distance $\r2$ divided by the  radius
of gyration $\rg2$ versus $n$.
For a Gaussian chain $\r2/6\rg2=1$.
(a) $T=448$~K and $\rho=0.766 \ {\rm g/cm}^3$. Data from model A for
bulk and RIS simulations.
(b) $T=400$~K and 
experimental density at $P\simeq 0.1$~MPa.  Data from models A, B, and C
for bulk simulations.
\label{f:rg}}
\end{figure}

The structural properties for the molecules as expressed by the
mean squared radius of gyration
$\rg2$, the two largest eigenvalues of the mass tensor $l_I^2$
and $l_2^2$, and the
mean squared end-to-end distance $\r2$ are collected in
Tables \ref{table3} and \ref{table4} for all the simulations.
Results for $\r2/6\rg2$ 
are shown in Fig.~\ref{f:rg}a for model A at $T=448$~K. Data  for  
both the EMD $(\rho=0.766\ {\rm g/cm}^3)$ 
and the RIS simulations
are presented.  Note that for small $n\alt 16$, this
ratio increases with increasing $n$.
For small $n$, the chains are relatively rigid.
As such, the change in chain dimensions
with increasing $n$ is similar to what can be calculated for a linear
string of uniform beads, where $\r2/6\rg2$ goes from $\simeq 1$ 
for  very small $n$
to $2$ for infinite $n$.  Above $n\agt 16$, flexibility of the chain becomes
significant and $\r2/6\rg2$ decreases with increasing $n$.
A similar result was found by Baschnagel {\it et al.} \cite{baschnagel92} and
Brown {\it et al.} \cite{brown9496}.
The EMD results show that, for the values of $n$ studied, the chains
are clearly extended and non-Gaussian as $\r2 > 6\rg2$. For
this model and temperature, the crossover to Gaussian
statistics occurs for  $n$ much greater than $100$ in agreement
with Baschnagel {\it et al.} \cite{baschnagel92},
Brown {\it et al.} \cite{brown9496}
and Paul {\it et al.} \cite{paul98}.  Figure \ref{f:rg}a shows that
the values of $\r2/6\rg2$ predicted by RIS
and EMD simulations agree within our statistical uncertainty.  For the
longest chain length studied, $n = 2000$, $\r2/6\rg2$ is roughly
$1\%$ greater than the prediction for Gaussian chains for model A.
A second order curve
has been constructed on Fig.~\ref{f:rg}a representing
the best fit of all the data for the RIS
simulations and the data from the EMD simulations at
$n = 66, 44, 36, 30, 24,$ and $16$.  For model A,
($\r2/6\rg2 = 1 + 11.8/n - 93.4/n^2$).  From this curve one can form an
approximation
of the chain length needed to reach Gaussian statistics
within  a given degree of error.
For instance, the deviation from Gaussian statistics is less
than $2\%$ for chain lengths $n\agt 600$ for model A.
This is considerably larger than for a flexible chain consisting
of beads connected
with a spring \cite{kremer90} but with 
no torsional or bending forces, in which $\r2/6\rg2=1$ for $n\ge 10$.

{From} our previous studies for small $n$ $(n\le 24)$, we
expect that the actual crossover to Gaussian chain structural
statistics will depend somewhat on the specific model, particularly
through the parameters of the torsional potential.
As discussed
in Sec.\ II, the
torsional potential
for model A has a larger
{\it gauche} bond energy than that used
in models B and C (about $0.4$ kcal/mol larger).
This generally means that
models B and C result in more thermodynamically
flexible, and therefore more compact, molecules than model A.
Data in Table \ref{table4} confirms these
expectations; furthermore, the ranking of $\r2$ for the three models
is constant with $n$:
$\r2_A > \r2_C > \r2_B$.  Comparing data for model A and model C,
the influence of the {\it gauche} bond energy on $\r2$ is evident.  This
comparison is valid as models A and C represent the same implementation
of the united atom model, only with different parameters.  Furthermore,
their largest difference in parameters, for this study of 
equilibrium properties, is found in their
torsional potentials.  Thus the effect on molecular size can be directly
attributed to the difference in {\it gauche} bond energy.  Comparing $\r2$
data for models B and C,
we note that the difference in molecular size for the two models
is larger than what one would expect due simply to the slight difference
in the {\it gauche} bond energy. 
This is sensible since model B represents
a fundamentally different parameterization of the united atom model than
models A and C.  This  indicates that  torsional potentials
are not directly transferable between UA and AUA models.
{From} observation of the data in Table \ref{table4}, we
anticipate that
the crossover to Gaussian structural statistics with increasing $n$
will occur first for
model B, then for model C, and last for model A; Fig.~\ref{f:rg}b
gives evidence of this.
This figure is the same as Fig.~\ref{f:rg}a but
with data from the EMD simulations using the three models
at $T=400$~K and
experimental density at atmospheric pressure.  While Fig.~\ref{f:rg}b
shows only the beginning of the crossover, the ranking predicted is
evident at this stage.  The RIS simulations for the three models
confirm the order of the crossover to Gaussian statistics.

Differences in the torsional potential also give
rise to a difference in the expansion coefficient
$k=d(\ln \rg2)/dT$ between the models.  In a previous publication
\cite{mondello95}, we found that for
tetracosane around room temperature,
$k=-1.1\pm 0.2$ and $-0.6\pm 0.3\times 10^{-3}$~K$^{-1}$
for model A and B, respectively.
These results were for simulations run at the experimental
density for $P=0.1$~MPa for each temperature.
This data was  compared
to the experimental value for bulk polyethylene
$k=-1.07\times 10^{-3}$~K$^{-1}$ obtained from
SANS measurements \cite{boothroyd91}.
We can also consider previous data for $\rg2$
of $n$-hexadecane \cite{mondello97} obtained at $T=298, 323,$ and
$373$~K
in conjunction with $\rg2$ obtained at $T=400$~K presented
in this paper.  From these, we can calculate
$k=-0.8\pm 0.2$ and $-0.6\pm 0.1\times 10^{-3}$~K$^{-1}$
for $n$-hexadecane in model A and B, respectively.
These results are not presented to further evaluate
the two models so much as to
demonstrate that there is a
very slow convergence of molecular
sizes for models A and B with increasing temperature. In this
respect, note that the average radius of gyration of 
$n$-hexadecane using model A
at $T=400$~K is still larger than the value for model B at
$298$~K \cite{mondello95}.  While temperature can influence
the position in $n$ of the crossover to Gaussian
statistics, these observations show that the model used has a
much greater influence than temperature.

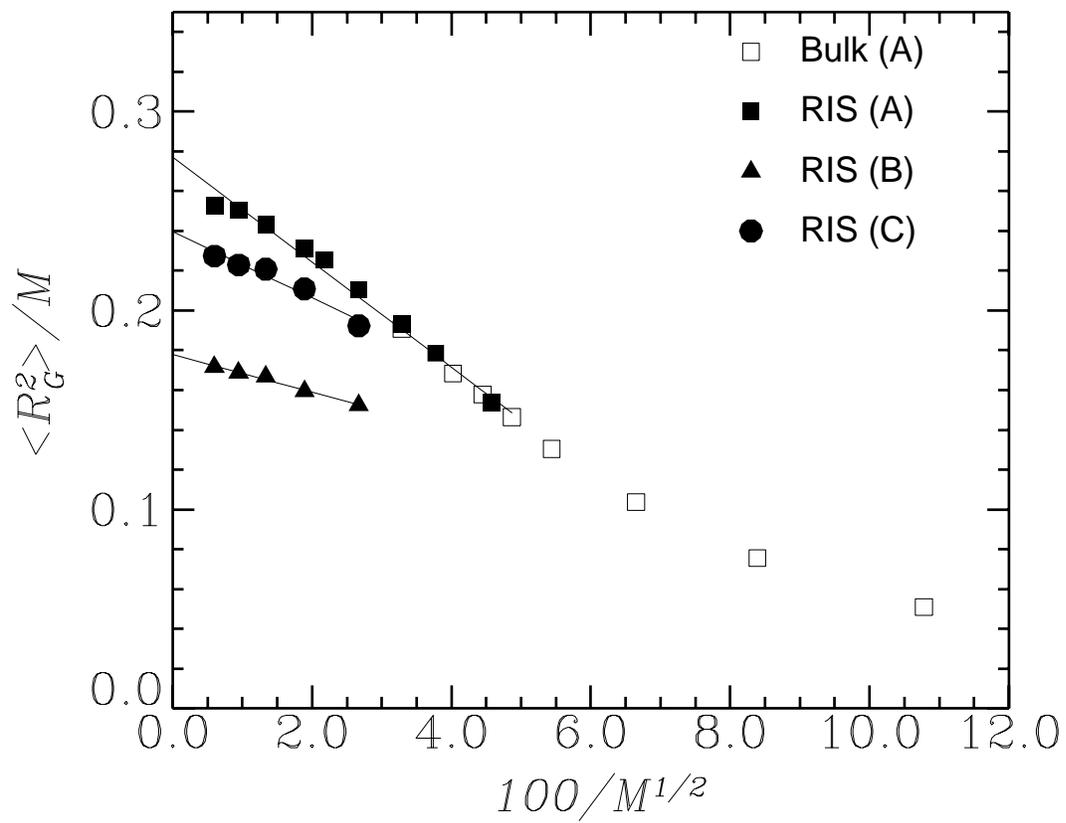
\begin{figure}[tb]
\caption{Radius of gyration $\rg2$ divided by mass M versus
$100 / M^{1/2}$ for $T=448$~K and $\rho=0.766 \ {\rm g/cm}^3$,
model A bulk and RIS simulations for models A, B, and C.
\label{f:rg21ov}}
\end{figure}

Boothroyd {\it et al.} \cite{boothroyd91} have reported 
on experimental data for the
radius of gyration  of a polyethylene chain in a melt.  
Specifically, they quote
the radius of gyration for samples  of molecular weight
$M_W=32000$ and $50000$ in the temperature
range of $380$~K to $480$~K.  Dividing the experimental $\rg2$
by $M_W$ for $M_W=32000$ and $50000$ at $448$~K, we
obtain $0.23$ and $0.19$, respectively.
In Fig.~\ref{f:rg21ov}, $\rg2/M_W$ 
is plotted versus $100/M_W^{1/2}$.  Data obtained
from both the EMD and RIS simulations at $T=448$~K are shown for
model A.  In addition, a linear fit to all the RIS data and the EMD
data for $30\le n \le 66$ is also shown.
Extrapolating this line to $M_W \rightarrow \infty$
gives a value of $0.28$ for model A which is considerably larger
than the  
experimental results for polyethylene. Similar fits to the RIS
model using the parameters for models B and C give $\rg2/M_W=0.18$ and
$0.24$, respectively, in better agreement with experiment.

\noindent {\bf IV. DIFFUSION AND VISCOSITY RESULTS}

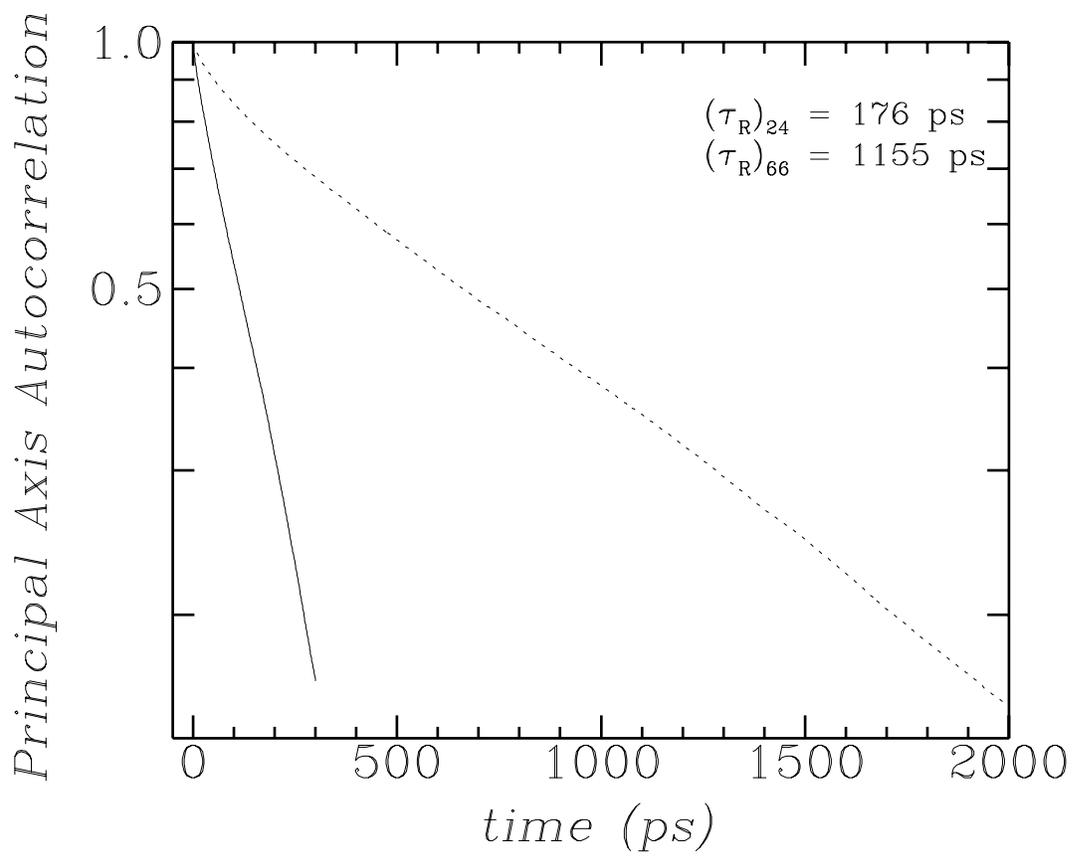
\begin{figure}[tb]
\caption{Time autocorrelation function of the  orientation of the
longest principal axis of inertia for $n$-tetracosane $(n=24)$ 
and $n$-hexahexacontane $(n=66)$ for $T=448$~K.  The solid line
is for $(n=24)$ and the dashed is for $(n=66)$.
\label{f:auto}}
\end{figure}

Results for the self-diffusion constant $D$, the rotational
diffusion time $\tau_R$ and the shear viscosity $\eta$ are
collected in Tables \ref{table3} and  \ref{table4} for all
simulations performed. 
The self-diffusion
constant
is obtained from the slope of the mean-squared displacement
of the molecular center of mass, averaged over all the molecules
$(N_{\rm mol})$ in the system and all available initial times.
In taking the average, the sum of the mean-squared displacements
for the $N_{\rm mol}$ is divided by $N_{\rm mol} -1$ to 
account for the fact that the center of mass of the system does not
move. To calculate $\tau_R$, we consider the first-order
angular correlation of the longest principal axis of a molecule's
ellipsoid of inertia \cite{mondello95}. At intermediate
times, this correlation is well described by a simple exponential
relaxation with a time constant $\tau_R$ as shown in Fig.\ 
\ref{f:auto} for $n$-tetracosane and  $n$-hexahexacontane.
Equivalently, one can measure the autocorrelation function
$\big<{\bf  R}(t)\cdot {\bf  R}(0)\big>$ for the end-to-end
distance \cite{paul95} and fit this curve with the corresponding
formula from
the Rouse model \cite{doi86} to obtain a measure of $\tau_R$.
Both of these correlations can be calculated from EMD. 
The viscosity was determined directly from NEMD
simulations.
{From} our previous study \cite{mondello97} comparing EMD and
NEMD simulation methods for determining the viscosity of 
$n$-alkanes, the NEMD simulation technique requires about
half the cpu time required by EMD methods for comparable
accuracy. For further details
see ref.~\cite{mondello97}.

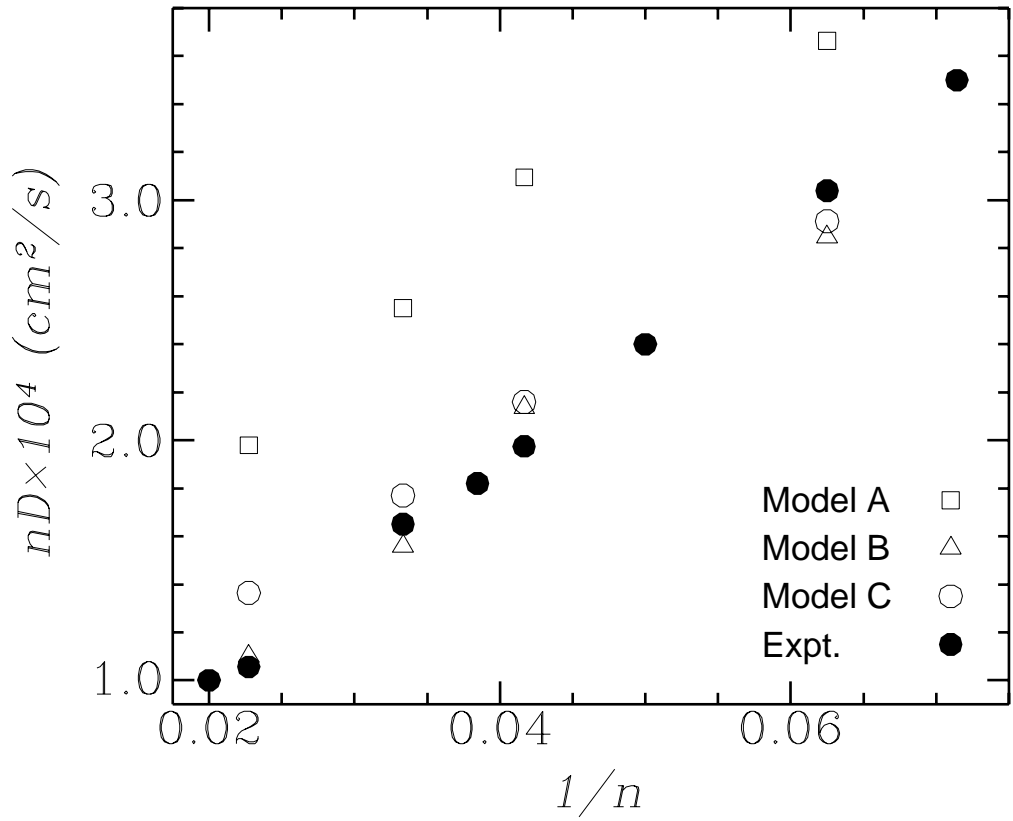
\begin{figure}[tb]
\caption{Diffusion constant $D$ from models A $(\Box)$,	
B $(\bigtriangleup)$, and C $(\circ)$ for $T=400$~K
at the experimental densities listed in Table \protect{\ref{table4}} for
$P=0.1$~MPa.  The experimental results $(\bullet)$
are from ref.~\protect{\cite{ednote2}}.
}
\label{f:D}
\end{figure}

The diffusion constants for the three models studied 
are compared to experimental data \cite{ednote2} for
$T=400$~K in Fig.~\ref{f:D}. These simulations were run
at the experimental densities listed in Table \ref{table4} for
$P=0.1$~MPa. Note that for all of the simulation results as well
as the experiments, the product $Dn$ decreases with increasing $n$. This 
decrease is due to the fact that the density and therefore the
monomeric friction coefficient $\zeta$ increases with increasing $n$,
causing $D$ at fixed $T$ to decrease faster than predicted by
the Rouse model. Assuming that  the change in $Dn$ is due to
change in the monomeric
friction coefficient a fit to the data  gives  $\zeta\sim n^{-0.8}$.
One also sees from Fig.~\ref{f:D} and Table \ref{table4} that model
A overestimates $D$ by approximately $15 - 50\%$, while models B and
C fit
the available experimental data quite well. 
Similar results for models A and B were found in our earlier studies
\cite{mondello95,mondello96,mondello97}
for $n$-decane, $n$-hexadecane, and $n$-tetracosane as well as for
several branched alkanes.  What is new in the current results is the
agreement with experiment found for model C which is a symmetric united
atom model.  A symmetric united atom model is somewhat less computationally
demanding to implement than an asymmetric united atom model.  For
identical length simulations
we find that for $100$ molecules of $n=44$,
the symmetric UA model requires approximately
$10\%$ less cpu time than the AUA model.

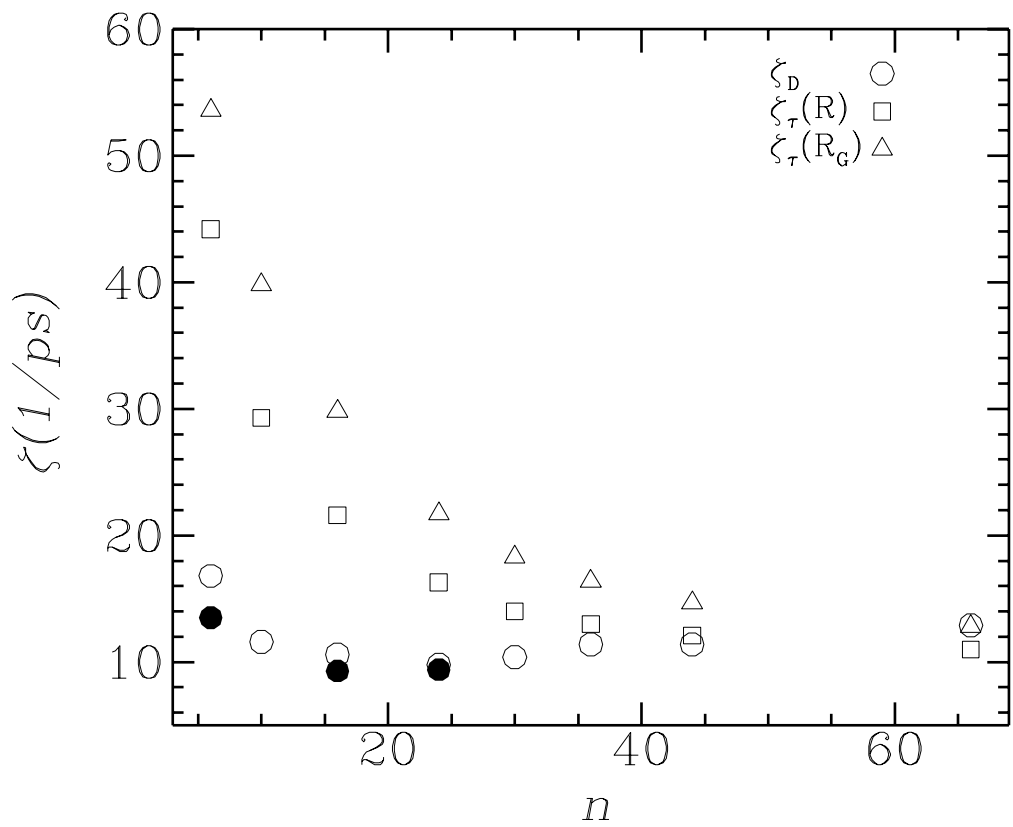
\begin{figure}[tb]
\caption{Friction coefficients $\zeta$ versus $n$ for $T=448$~K.  
$\zeta_D \ (\circ)$
is extracted from the self-diffusion constant, eq.~\protect\ref{eq:zetaD},
while $\zeta_\tau(R)\ (\Box)$  and $\zeta_\tau(R_G)\ (\bigtriangleup)$ 
are determined from the
longest relaxation time $\tau_R$, eq.~\protect\ref{eq:zetatau}.
The filled symbol $(\bullet)$ is $\zeta_D$ obtained from simulations
for chains consisting entirely of ${\rm CH}_2$ united atoms.
\label{f:zeta}}
\end{figure}

The three estimates of $\zeta$ are shown in Fig.~\ref{f:zeta} for
constant density and $T=448$~K. Note that $\zeta_D$ determined
from the diffusion constant is only weakly dependent on $n$ and
even very small chain lengths give a reasonably good estimate of the
friction coefficient of unentangled, Rouse-like chains. The change
in $\zeta_D$ from $n=10$ to $66$ is only $10\%$. 
The significant change in $\zeta$ with $n$ at constant pressure
as exhibited by the diffusion
data discussed above all
but disappears at constant density.
This is in contrast
with $\zeta_\tau$ which depends much more strongly on $n$ for small $n$;
furthermore, $\zeta_\tau >\zeta_D$ \cite{paul95}.  The difference between
$\zeta_\tau(R_G)$ and $\zeta_\tau(R)$ for small $n$ is another
representation of the deviation of equilibrium structural statistics
from Rouse-like behavior.
For $n\agt 60$, all three methods of determining $\zeta$ agree within
our statistical uncertainties as predicted by the Rouse model. 

Another attribute of Fig.~\ref{f:zeta} warrants discussion.  At small $n$,
$\zeta_D$ decreases with increasing $n$ and appears to reach a minimum
around $n=24$.
This behavior is caused by an
end effect:  at small $n$, there is a large concentration of ${\rm CH}_3$ united
atoms whose Lennard-Jones $\epsilon$ parameter in model A is over twice that
for ${\rm CH}_2$ united atoms.  An increase in $\epsilon$ is known to decrease
diffusion \cite{mondello95} thereby effectively increasing $\zeta_D$.
To verify that this end effect caused the increase in $\zeta_D$, the state
points for $n=6, 16,$ and $24$ were simulated
but with the $\epsilon=0.93$ kcal/mol for all the united atoms
(i.e.\ the value for ${\rm CH}_2$ in Model A).  The mass of all united atom
groups was that of ${\rm CH}_2$ so the volumes were adjusted to achieve the
stated density.
The results for $\zeta_D$ are shown on
Fig.~\ref{f:zeta} as filled symbols.  The change in value at $n = 24$ is very
small ($4\%$) demonstrating that the end effect is nearly undetectable for    
large $n$ as expected.  
The overall variation in $\zeta_D$ is now also smaller.  From
the three data points presented, it can be seen that the minimum in $\zeta_D$
with $n$ still exists, however the minimum has shifted to $n = 16$,
coincident with the position of the maximum in $\rg2$ (Fig.~\ref{f:rg}a).  
This suggests that the origin of the minimum of $\zeta_d$ in Fig.~\ref{f:zeta}
is related to the chain flexibility.

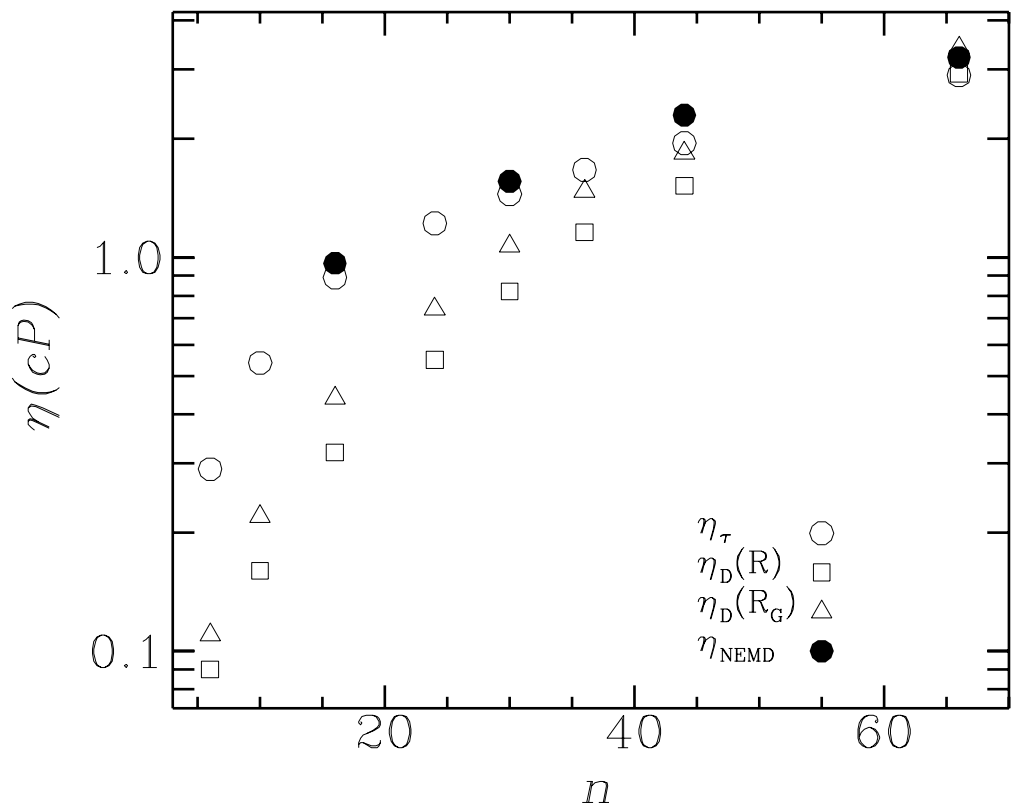
\begin{figure}[tb]
\caption{Viscosity $\eta_\tau\ (\circ)$  
determined from the longest relaxation time
$\tau_R$, eq.~\protect{\ref{eq:vistau}}, and 
$\eta_D(R)\ (\Box)$ and $\eta_D(R_G)\ (\bigtriangleup)$
determined from the self-diffusion constant
$D$, eq.~\protect{\ref{eq:visD}}, versus $n$ for $T=448$~K. 
Also shown for four values of $n$ is the viscosity $\eta\ (\bullet)$ 
measured in our non-equilibrium molecular dynamics
simulations.  
\label{f:vis}}
\end{figure}

Finally, the three estimates for the viscosity are compared to
our results from NEMD simulations in Fig.~\ref{f:vis}. As observed
in our previous study \cite{mondello97}, $\eta_\tau$ gives a very
good estimate (within about $20\%$) for all $n$, while $\eta_D$ is
significantly smaller for small $n$. Since $\tau_R$ requires     
approximately ten times less cpu time to determine than $\eta$, we
believe that at least for unentangled chains, 
$\tau_R$ is an efficient way to predict $\eta$ without having
to measure $\eta$ directly.  For $n\agt 60$, all the estimates of
$\eta$ and the NEMD calculation agree within our statistical
uncertainty.

\noindent {\bf V. CONCLUSIONS}

The crossover to Rouse-like behavior  for $n$-alkanes $(6 \le n \le 66)$
as demonstrated by their self-diffusion 
constant $D$, viscosity $\eta$, and equilibrium structural
statistics
was studied numerically. 
For small $n$ the chains were non-Gaussian and the mean
squared end-to-end distance $\r2$ was greater than $6\rg2$, where
$\rg2$ is the mean squared radius of gyration. As $n$ increases, $\r2/\rg2 
\rightarrow 6$ but only for $n$
significantly larger than $200$.
At constant density, the Rouse model was used to extract
the monomeric friction coefficient 
$\zeta$ and the viscosity $\eta$ independently from the diffusion
constant $D$ and the
longest relaxation time $\tau_R$.  $\zeta_D$ extracted
from $D$ was nearly independent of chain length while 
$\zeta_\tau$ obtained from
$\tau_R$ was much larger than $\zeta_D$ for small $n$.
The viscosity measured in a non-equilibrium
molecular dynamics simulation was closely approximated by
the value of $\eta$ determined from $\tau_R$ while 
the $\eta$ inferred  from $D$ is smaller for small $n$. 
For $n\agt 60$, the
two estimates for both  $\zeta$ and $\eta$ agree
as predicted from the Rouse model.
Diffusion at normal pressure as calculated from three interaction models
was studied for increasing $n$
and the results were compared
to experimental data.  Very good agreement with
experiment was found for two of the models while the third consistently
overestimated $D$.  For all models and experimental data, $D$ was found
to depend more strongly on $n$ than predicted by Rouse theory.


\newpage

\newpage

\begin{table}
\caption{Intramolecular interaction parameters.}
\begin{tabular}{d d d d d}
    & Model A (UA\tablenotemark[1]) & Model B (AUA\tablenotemark[2]) & Model C (UA\tablenotemark[3]) & Units  \\
\hline
bond length & 1.54 & 1.54 & 1.54 & $\AA$\\
k$_{b}$ (bending)& 124.18 & 124.28 &  124.18 &  kcal/(mol rad$^2$)  \\
 $\theta_{b}$ & 114.0$^o$ & 114.6$^o$ & 114.0$^o$ & \\
a$_{0}$ (torsion) & 2.007 &  2.062 &  1.736 &  kcal/mol \\
a$_{1}$ & 4.012 &  4.821 &  4.500 &            \\
a$_{2}$ & 0.271 &  0.162 &  0.764 &    \\
a$_{3}$ & -6.290& -6.218 & -7.000 &    \\
a$_{4}$ &       & -0.324 &        &    \\
a$_{5}$ &       & -0.502 &        &    \\
\end{tabular}
\tablenotetext[1]{Intramolecular parameters for n-alkanes from
Ref.~\onlinecite{siepmann93,smit95}. Torsional potentials are taken from
Ref.~\cite{jorgensen84}.
}
\tablenotetext[2]{Intramolecular parameters from
Ref.~\onlinecite{padilla91}; torsional potential (d) was used.
}
\tablenotetext[3]{The torsional potential is the ``modified" one from
Ref.~\onlinecite{paul95}.  The bending term used here is slightly
different from the one used by Paul {\it et al}; see \cite{ednote3}.
}
\label{table1}
\end{table}

\begin{table}
\caption{Lennard-Jones potential parameters.}
\begin{tabular}{d d d d d}
Model & Group & $\sigma$ $(\AA)$& $\epsilon$ (kcal/mol) & $d$ $(\AA)$\\ \hline
A (UA\tablenotemark[1])   & CH$_{3}$ &  3.930&  0.227 & \\
         & CH$_{2}$ &  3.930&  0.093 & \\
B (AUA\tablenotemark[2])  & CH$_{3}$ &  3.527&  0.238 & 0.275 \\
         & CH$_{2}$ &  3.527&  0.159 & 0.370 \\
C (UA\tablenotemark[3])   & CH$_{3}$ &  4.010&  0.227 & \\
         & CH$_{2}$ &  4.010&  0.093 & \\
\end{tabular}
\tablenotetext[1]{From Ref.~\onlinecite{siepmann93,smit95}.
}
\tablenotetext[2]{From Ref.~\onlinecite{padilla91}.
}
\tablenotetext[3]{The value used in Ref.~\onlinecite{paul95} 
for $\epsilon_{CH_3}$ was $0.226$ kcal/mol and 
the $\sigma$ for both ${\rm CH}_3$ and ${\rm CH}_2$ united atoms
was $4.009 \AA$.  
}
\label{table2}
\end{table}

\begin{table}
\caption{Summary of the constant density ($\rho = 0.766$ g/cm$^3$) simulation
results for model A at $T=448$~K.  Results for the mean squared
radius of gyration $\rg2$ and the end-to-end distance $\r2$ are expressed in
$\AA^2$. The eigenvalues of the mass tensor $(l^2_I)$ satisfy
the equality $\rg2=l^2_1+l^2_2+l^2_3$, with $l^2_1>l^2_2>l^2_3$. 
Results for the self-diffusion and rotational relaxation time $\tau_R$
are from equilibrium molecular dynamics simulations, while the results
for the viscosity $\eta$ are from non-equilibrium molecular dynamics
simulations.
The total length of the run and the
number of molecules used are also shown.  
Uncertainties in the last reported digit(s) are given in parenthesis.   
}
\begin{tabular}{ccccdccdddc}
       & $n$ & $N$ & $t$ (ns) & $\rg2$ & l$_{1}^2$/$\rg2$ & l$_{2}^2$/$\rg2$ &  $\r2$  & $D$ ($10^{-6} {\rm cm}^2/$sec) & $\tau_R$ (ps)\\ \hline
EMD    &  6  & 64  &  1.0  &    4.39(1)  & 0.897(1)  &  0.082(1)  & 31.9(1) & 25.8(7) &   11 & \\
       & 10  & 64  &  1.5  &    10.7(1)  & 0.900(1)  &  0.082(1)  & 87.5(2) & 22.7(7) &   33 & \\
       & 16  & 64  &  1.5  &    23.5(1)  & 0.876(1)  &  0.105(1)  &  194(1) & 15.5(7) &   86 & \\
       & 24  & 100 &  1.0  &    44.1(2)  & 0.847(2)  &  0.130(2)  &  353(3) & 11.2(5) &  176 & \\
       & 30  & 100 &  1.5  &    61.8(4)  & 0.833(3)  &  0.140(2)  &  484(6) & 8.56(31)&  260 & \\
       & 36  & 100 &  2.0  &    79.8(7)  & 0.821(3)  &  0.148(3)  &  605(10)& 6.40(34)&  360 & \\
       & 44  & 100 &  2.0  &     104(1)  & 0.804(4)  &  0.159(3)  &  758(16)& 5.27(30)&  513 & \\
       & 66  & 100 & 14.0  &     178(1)  & 0.785(2)  &  0.170(2)  & 1240(17)& 3.12(8) & 1140 & \\
       &     &     &        &           &            &         &         &    & \\
       & $n$ & $N$ &$t$ (ns)&$\eta$ (cP)&            &         &         &    & \\
NEMD   &  16 & 100 &  18.5  & 0.967(70) &            &         &         &    & \\
       &  30 &  64 &  18.0  & 1.56(6)   &            &         &         &    & \\
       &  44 & 100 &  13.5  & 2.30(5)   &            &         &         &    & \\
       &  66 & 100 &  10.1  & 3.22(5)   &            &         &         &    & \\
\end{tabular}
\label{table3}
\end{table}

\begin{table}
{
\renewcommand \baselinestretch 1
\caption{
Summary of the simulations at the experimental density
for $P\simeq 0.1$~MPa and $T=400$~K. 
For each chain length, data is presented from simulations employing models
A, B, and C.
Densities $\rho$ are quoted in gm/cm$^3$.
The results for the mean squared
radius of gyration $\rg2$ and the end-to-end distance $\r2$ are expressed in
$\AA^2$. The eigenvalues of the mass tensor $(l^2_I)$ satisfy
the equality $\rg2=l^2_1+l^2_2+l^2_3$, with $l^2_1>l^2_2>l^2_3$. 
The total length of the run and the
number of molecules used are also shown.  
Uncertainties in the last reported digit(s) are given in parenthesis.   
\label{table4}
}
}
\begin{tabular}{ccccdccddd}
Substance & Model & $N$ & $t$ (ns) & $\rg2$ & l$_{1}^2$/$\rg2$ & l$_{2}^2$/$\rg2$ & $\r2$ & $D$ & $D_{exp}$ \\ \hline
{\it n}-$C_{16}$ & A & 64 & 2.0 & 24.1(1) & 0.881(1) & 0.102(1) & 201(1) & 23.2(1.3) & 20(1) \\
$\rho=0.6999$\tablenotemark[1]    & B & 64 & 2.0 & 21.8(1) & 0.858(1) & 0.119(1) & 173(1) &18.1(9)& \\
                 & C & 64 & 2.0 & 23.3(1) & 0.876(1) & 0.105(1) & 192(1) &18.5(1.0)& \\
       & & & & & & & & & \\
{\it n}-$C_{24}$ & A & 100 & 1.0 & 45.3(3) & 0.852(3) &          & 367(5) &12.9(9)& 8.23(82) \\
$\rho=0.7260$\tablenotemark[2]    & B & 100 & 1.0 & 40.1(4) & 0.833(4) &          & 309(6) &8.90(62)& \\
                 & C & 100 & 2.0 & 43.3(2) & 0.845(2) & 0.131(2) & 344(3) &9.11(28)& \\
       & & & & & & & & & \\
{\it n}-$C_{30}$ & A &  64 & 4.0 & 64.0(4) & 0.838(2) & 0.137(2) & 507(5) &8.62(47)& 5.2(3) \\
$\rho=0.7421$\tablenotemark[3]    & B &  64 & 4.0 & 55.3(4) & 0.820(3) & 0.148(2) & 415(5) &5.30(30)& \\
                 & C &  64 & 4.0 & 61.0(4) & 0.833(3) & 0.140(2) & 477(6) &6.03(34)& \\
       & & & & & & & & & \\
{\it n}-$C_{44}$ & A & 100 & 4.0 &  111(1) & 0.815(3) & 0.152(3) & 833(15)&4.52(21)& 2.4(2) \\
$\rho=0.7570$\tablenotemark[4]    & B & 100 & 4.0 &92.3(1.2) & 0.794(4) & 0.165(3) & 654(15)&2.49(12)& \\
                 & C & 100 & 4.0 &  104(1) & 0.807(4) & 0.157(3) & 765(16)&3.17(17)& \\
       & & & & & & & & & \\
\end{tabular}
\tablenotetext[1]{Experimental densities at ten temperatures
were obtained from Ref.~\onlinecite{dymond79,dymond80,API42} and used to
create a linear extrapolation to $T=400$~K.
}
\tablenotetext[2]{From Ref.~\onlinecite{Nederbragt47}.
}
\tablenotetext[3]{Obtained by a linear interpolation between experimental
densities from Ref.~\onlinecite{doolittle64}.
}
\tablenotetext[4]{Calculated based on information from
Ref.~\onlinecite{paul95}.  While
no experimental reference is given in Ref.~\onlinecite{paul95},
a comparison of the value used was made to the value obtained by a
linear interpolation between data from Ref.~\onlinecite{API42}.  The
two values differed by $<1\%$.
}
\end{table}
\end{document}